\documentclass[conference]{IEEEtran}
\IEEEoverridecommandlockouts

\usepackage{amssymb}
\usepackage{multirow}
\usepackage[caption=false]{subfig}
\usepackage{cite}
\usepackage{amsmath,amssymb,amsfonts}
\usepackage{algorithmic}
\usepackage{graphicx}
\usepackage{textcomp}
\usepackage{xcolor}
\def\BibTeX{{\rm B\kern-.05em{\sc i\kern-.025em b}\kern-.08em
    T\kern-.1667em\lower.7ex\hbox{E}\kern-.125emX}}
\begin{document}

\title{Speech Emotion Recognition with Phonation Excitation Information and Articulatory Kinematics}

\author{\IEEEauthorblockN{Ziqian Zhang, Min Huang, Zhongzhe Xiao}
	\IEEEauthorblockA{School of Optoelectronic Science and Engineering\\
		Soochow University, Suzhou, China}}

\maketitle

\begin{abstract}
Speech emotion recognition (SER) has advanced significantly for the sake of deep-learning methods, while textual information further enhances its performance. However, few studies have focused on the physiological information during speech production, which also encompasses speaker traits, including emotional states. To bridge this gap, we conducted a series of experiments to investigate the potential of the phonation excitation information and articulatory kinematics for SER. Due to the scarcity of training data for this purpose, we introduce a portrayed emotional dataset, STEM-E$^\textbf{2}$VA, which includes audio and physiological data such as electroglottography (EGG) and electromagnetic articulography (EMA). EGG and EMA provide information of phonation excitation and articulatory kinematics, respectively. Additionally, we performed emotion recognition using estimated physiological data derived through inversion methods from speech, instead of collected EGG and EMA, to explore the feasibility of applying such physiological information in real-world SER. Experimental results confirm the effectiveness of incorporating physiological information about speech production for SER and demonstrate its potential for practical use in real-world scenarios.
\end{abstract}

\begin{IEEEkeywords}
Speech emotion recognition; speech production; excitation source; articulatory kinematics
\end{IEEEkeywords}

\section{INTRODUCTION}
Speech emotion recognition (SER) is a crucial issue that aims to understand and identify the emotional states of human from speech. It plays an essential role in advanced technologies, such as human-computer interaction, human–robot interaction, and physiological conditions detection \cite{de2023ongoing}. As the rapid advancements in natural language processing (NLP), computer vision (CV), and speech analysis, tremendous attempts have been made to improve SER performance from various perspectives \cite{akinpelu2024enhanced,Wang2024_icassp,Chandola2024_icassp,pentari2024speech,niu2024_interspeech,santoso2024ICASSP,Ma2024_icassp,singh2021multimodal,He2024_icassp}. Many studies have achieved state-of-the-art results in SER with deep-learning-based models such as Transformer and Graph Convolutional Network (GCN)  \cite{akinpelu2024enhanced, Wang2024_icassp,Chandola2024_icassp,pentari2024speech}. Some research focuses on annotating emotional speeches utilizing Large Language Models (LLMs) to mitigate the subjectivity and ambiguity in human annotations \cite{niu2024_interspeech,santoso2024ICASSP}. Another growing area of interest is the use of speech pre-trained models to boost SER \cite{Ma2024_icassp}. Additionally, integrating audio with textual information in speech has also been explored to more comprehensively identify human emotions \cite{singh2021multimodal,He2024_icassp}. Despite the progress made, there remains a necessity to deepen our understanding of the physiological information underlying speech production, as speech can be viewed as the output of an excitation filtered by the vocal tract.

Speech is produced by the physiological apparatus of the human speech production system, which consists mainly of two components: phonation excitation, primarily generated in the larynx, and filtering, which refers to the effects of the vocal tract on excitation during speech production \cite{fant1971acoustic}. Phonation excitation is the air flow waveform generated by the vibration of vocal folds, while the filtering can be influenced by the condition of articulators, such as the positioning of the tongue, the opening degree of the mouth, and the movement of the lips \cite{kadiri2021extraction}. Hence, variations in both excitation and articulators will change the acoustical cues of speech that convey speaker traits, including the emotional states \cite{Sundberg2011Interdependencies,lee05c_interspeech}.  Based on this, studies leveraging excitation information or articulatory movements have shown their auxiliary effectiveness in aiding SER \cite{xiao2018contribution,chen2013speech,ren2018articulatory,Li2024speech}.

However, few studies that have investigated the effectiveness of excitation information or articulatory movements for SER, particularly in their combined ability to distinguish between different emotional states. One reason for this gap is the limited availability of data on the physiological process involved in speech production, such as electroglottography (EGG), electromagnetic articulography (EMA), and magnetic resonance imaging (MRI). Additionally, the complexity and high cost of collecting such physiological data make it impractical for real-world SER systems.

In this paper, we hypothesize that speech production-related physiological information can enhance the performance of SER, especially when the information of phonation excitation and ariculatory kinematics are included. To verify this hypothesis, we perform emotion recognition by fusing speech data with excitation information and articulatory movements using our collected Mandarin emotional dataset, "Suzhou \& Taiyuan Emotional dataset on Mandarin - EMA, EGG, Video, Audio" (STEM-$\rm{E^2}$VA). The excitation information is represented by EGG, which provides observation of vocal fold vibratory activity during voice production \cite{herbst2020}. Articulatory movements are recorded using an EMA system, tracking the motion of the tongue and lips. This study employs a conventional emotion recognition architecture, where features are manually extracted and input into a machine learning model. In addition, motivated by inversion methods that estimate excitation information \cite{prathosh2019adversarial,kadiri2021extraction} and articulatory movements \cite{liu2015,shahrebabaki2021,cho2024_icassp} from speech without specialized equipment, we also conduct an initial investigation into the effectiveness of predicted physiological data for SER. 

\section{METHODOLOGY}

\subsection{Collection of STEM-${E^2}$VA dataset}
To validate the effectiveness of excitation information and articulatory movements in emotion recognition, we recorded parallel audio, EGG, and EMA data, together with video resulting in STEM-${E^2}$VA dataset. 24 native Mandarin Chinese speakers (12 females, 12 males), all university students aged 23 to 28, were selected for this recording. Prior to recording, participants were asked to complete the Symptom Checklist-90 (SCL-90), a 90-item questionnaire used to assess psychological problems, to ensure they had normal emotional expression and perception abilities. Only those who passed the test and had no vocal disorders were included in the recording. During the session, speakers were asked to produce 16 scripted sentences \cite{xiao2018contribution} with seven emotions: neutral, ecstatic, pleased, angry, indifferent, pained, and sad, where the intensity states of emotions were considered, resulting in 112 utterances per speaker (16 sentences × 7 emotions).

\begin{figure}
	\centering
	\includegraphics[width=.8\linewidth]{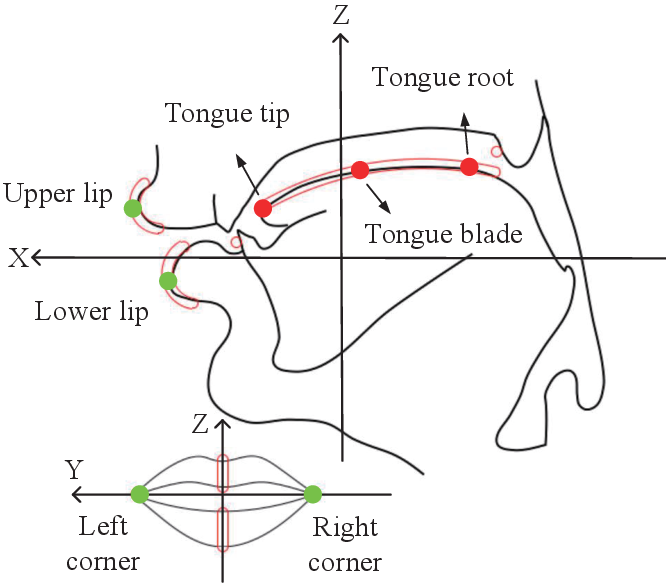}
	\caption{Configuration of sensors in the EMA system.}
	\label{Fig:sensor_configuration}
\end{figure}

Parallel audio and EMA data were recorded using the Carstens AG501 electromagnetic articulography system, which tracks the movement of sensors attached to key articulators. The positions of seven sensors were selected for data collection: four sensors were attached around the lips (upper lip center (UL), lower lip center (LL), left lip corner (LC) and right lip corner (RC)), and three sensors were placed along the midsagittal plane of the tongue (tongue tip (TT), tongue blade (TB) and tongue root (TR)) , as shown in Fig.~\ref{Fig:sensor_configuration}. The positions of each sensor are described in six-dimensional (6D) data, which include x, y, z coordinates and rotations around the x, y, z axes (degree) with respect to the reference dimensional space of the EMA system. In this system, the x-axis corresponds to forward-backward movement, the y-axis corresponds to left-right movement, and the z-axis corresponds to upward-downward movement. The audio data collected are sampled at 48 kHz, while the EMA data have a sampling rate of 250 Hz. 

EGG data were collected using the D100 electroglottography to measure changes in relative vocal fold contact area during laryngeal voice production. Two electrodes were placed on each side of the thyroid cartilage of speakers, and a high-frequency, low-amperage current was passed between them. Changes in vocal fold contact area during voicing were captured as EGG signals, with a sampling frequency of 44.1 kHz. 

After recording, samples of abnormalities were manually reviewed and removed, such as instances of sensor data loss caused by recording errors and audio samples with excessive background noise. After this filtering process, a total of 2427 parallel audio, EMA and EGG samples were obtained, featuring data from 22 speakers (11 females, 11 males).

\subsection{Feature extraction}
In order to extract salient features from audio for emotion recognition, the standard feature set in the INTERSPEECH 2009 Emotion Challenge \cite{schuller2009} is utilized. This set is composed of 384 features, derived by applying 12 statistical functions (referred to as “functionals” in the INTERSPEECH challenge) to 16 low-level descriptors (LLDs) and their delta coefficients. Although larger feature sets have been introduced in subsequent INTERSPEECH challenges \cite{schuller2013}, this smaller set is more suitable to our small-scale dataset, which contains only 2427 samples. The same feature set was also applied to the 1-dimensional EGG data. The feature extraction for both audio and EGG was implemented using the openSMILE toolkit \cite{eyben2010opensmile}, with a frame length of 20 ms and a frameshift of 10 ms.

To minimize the influence of absolute positions of articuators, this study extracted velocity and accelerate features from EMA data by calculating the first and second derivatives, respectively, of the 21-dimensional (7 articulators $\times$ 3 coordinates) position data. Four Statistical functions were then applied to these values in a sentence-based manner: mean, maximum, variance, and standard deviation. Additionally, features extracted from vocal tract variables (VTVs), which rely on the coordination of multiple articulators, were included for emotion analysis. Based on articulatory phonology \cite{browman1992}, VTVs define the kinematic state of five distinct constrictors (lips, tongue tip, tongue body, velum, and glottis) located along the vocal tract in terms of their constriction degree and location, independent of the absolute position of the articulators \cite{ghosh2011subject}. Due to limitations in the articulators captured by our EMA data, 5 VTVs were selected for this study: lip aperture (LA), lip protrusion (LP), tongue tip constriction location (TTCL), tongue body constriction location (TBCL) and tongue root constriction location (TRCL). 

Following the method in \cite{seneviratne2019multi}, LA is defined as the Euclidean distance between the lower lip (LL) and upper lip (UL), and we modified the calculation of it as follows:
\begin{equation}
	\text{LA(n)}=\sqrt{X^2+Y^2+Z^2}
	\label{equ1}
\end{equation}
where X, Y, Z are the coordinates calculated by the following equations:
\begin{equation}
	\text{X}=LL_x(n)-UL_x(n)
	\label{equ2}
\end{equation}
\begin{equation}
	\text{Y}=LL_y(n)-UL_y(n)
	\label{equ3}
\end{equation}
\begin{equation}
	\text{Z}=LL_z(n)-UL_z(n)
	\label{equ4}
\end{equation}

LP is defined as the displacement along the x-axis of the LL sensor from its median position across all utterances $U$ for that speaker, as shown in (\ref{equ5}).
\begin{equation}
	\text{LP(n)}=LL_x(n)-medium_{m\in{U}}(LL_x(m))
	\label{equ5}
\end{equation}

The Tongue Constriction Locations (TTCL, TBCL and TRCL) are also defined as the displacement of the sensor along the x-direction from their median position across all utterances $U$ for that speaker, as shown in (\ref{equ6}) to (\ref{equ8}).
\begin{equation}
	\text{TTCL}=medium_{m\in{U}}(TT_x(m))-TT_x(n)
	\label{equ6}
\end{equation}
\begin{equation}
	\text{TBCL}=medium_{m\in{U}}(TB_x(m))-TB_x(n)
	\label{equ7}
\end{equation}
\begin{equation}
	\text{TRCL}=medium_{m\in{U}}(TR_x(m))-TR_x(n)
	\label{equ8}
\end{equation}

VTVs are calculated at each EMA sampling point. The final VTVs-based features for emotion recognition are derived by applying statistical functions to the VTVs values across all points within the sentence. A total of 29 statistical features are defined, including the mean, maximum, range, variance and standard deviation for all five VTVs, and the minimum for four VTVs (excluding LA). As a result, the feature vector extracted from EMA data per sentence contains 197 features (84 velocity features, 84 accelerate features, and 29 VTVs-based features).

\subsection{Recognition model}
Given that STEM-$\rm{E^2}$VA is relatively small, we selected XGBoost for this work, as it is a gradient boosting algorithm trees known to perform well on smaller datasets \cite{liang2020predicting}. XGBoost is an advanced implementation of the gradient boosting framework \cite{chen2016xgboost}. 
Like other boosting methods, it builds an ensemble of models in a sequential manner, where each new model is added to correct the errors of the previous models.

\subsection{Estimation of excitation information and articulatory movements}
The requirement of specialized equipment to collect EMA and EGG data makes emotion recognition systems less feasible in the real-world applications. To explore the potential of using speech production-related physiological information in practical settings, we additionally performed emotion recognition by leveraging estimated excitation information and articulatory movements derived directly from speech signals.

As an alternative to EGG that directly measuring vocal fold activity, this study estimated the glottal flow waveform from speech to provide excitation information using iterative adaptive inverse filtering (IAIF) \cite{alku1992glottal}. Articulatory movements of lips and tongue, which are typically captured by EMA, were estimated using acoustic-to-articulatory inversion (AAI) methods from more accessible speech data. We explored the use of HuBERT self-supervised learning features\cite{hsu2021hubert} as input acoustic features, which has obtained the state-of-the-art performance in current AAI systems of normal speech \cite{cho2023evidence, attia2023improving, wu2023speaker}, for our AAI system on emotional speech. The HuBERT-large model, pre-trained on the Librilight dataset (60,000 hours) was employed for feature extraction. For the acoustic-to-articulatory inversion, we implemented a Temporal Convolutional Network (TCN) following the approach in \cite{siriwardena2023secret}. The model was tasked with predicting the movements of seven articulators (in the x, y, and z directions) as well as five VTVs features: lip aperture (LA), lip protrusion (PRO), tongue tip constriction location (TTCL)tongue body constriction location (TBCL), and tongue root constriction location (TRCL).

\section{EXPERIMENTS}
To describe the experimental settings and results more clearly, in the following sections, we refer to the three types of input---speech, excitation information, and articulatory movements---as three modalities.

\subsection{\label{subsec:3:1} Experimental settings}
To test both ground-truth and estimated data, we defined three 7-class recognition scenarios based on the hypothesis that incorporating more information in the input will improve recognition accuracy: (1) \textbf{Uni-modal:} In this baseline scenario, only one modality was used---either speech, excitation information or articulatory movements. (2) \textbf{Bi-modal:} In this scenario, pairwise combinations of the three modalities were evaluated to discover that to what extent the bi-modal data can cover the necessary information needed in emotional voicing. (3) \textbf{Tri-modal:} In the last scenario, all three available modalities were fused together to maximize the emotional information collected. In scenarios (2) and (3), the features from each modality are combined with early fusion, where they are concatenated directly.

To ensure that the results from each scenario are comparable and reliable, all experiments were conducted using the same algorithm, namely XGBoost, under identical hyperparameter settings. To mitigate the influence of potential variations in sample distributions within the dataset, the recognition experiments were conducted in a 10-fold cross-validation manner.

\subsection{\label{subsec:3:2} Experimental results}
\subsubsection{Emotion recognition based on the ground-truth data}
The recognition accuracies according to the experimental settings are listed in TABLE~\ref{tab:results}. Uni-modal cases exhibit the lowest accuracies, due to the incomplete presentation of phonation information, bi-modal and tri-modal cases obtain significant improvement in comparison to uni-modal cases.

\begin{table}
	\caption{Comparison result of uni-modal, bi-modal, tri-modal emotion recognition based on ground-truth data. '\checkmark' means which modality is used. \label{tab:results}}
	\begin{center}
		\begin{tabular}{c|c|c|c}
			\hline
			\multirow{2}{*}{\textbf{Speech}} & \textbf{Excitation} & \textbf{Articulatory} & \multirow{2}{*}{\textbf{Accuracy}(\%)}\\
			 & \textbf{information} & \textbf{movements}\\
			\hline
			\checkmark & & & 79.97\\
			 & \checkmark & & 76.68\\
			 & & \checkmark & 76.59\\
			\checkmark & \checkmark & & 83.81\\
			\checkmark & & \checkmark & 87.23\\
			 & \checkmark & \checkmark & 85.17\\
			\checkmark & \checkmark & \checkmark & \textbf{88.42}\\
			\hline
		\end{tabular}
	\end{center}
\end{table}

In mono-modal recognition tasks, audio, which represents the response of the excitation source through the vocal tract filter, yielded the highest accuracy among 3 individual modalities. The two physiological modalities can only carry a part of cues in emotional speech production, thus the recognition accuracies obtained from these modalities were lower than the whole speech signals. In bi-modal scenario, all 3 combinations of two modalities achieved better recognition than the best uni-modal result. This indicates that a part of excitation or vocal tract information is lost during this interaction between excitation and vocal tract, thus the combination of excitation information and articulatory movements with speech can both improve the emotion expressing ability than mono-speech. The combination of excitation information and articulatory movements, which directly reflects the pronouncing physiology of vocal folds and vocal tract respectively, also presented much higher accuracy than any of the three mono modalities. This indicates that the interaction process between excitation source and vocal tract itself also provide a small amount of information in emotion expressing. When all three modalities were fully considered together to form the tri-modal scenario, the recognition accuracy was further increased than the bi-modal cases, with improvement of $1.19\%$ to the best bi-modal accuracy, and $8.45\%$ to the best uni-modal accuracy. This again proves that each modality can provide a part of unique emotional information which is not contained in other modalities.

\begin{figure}[htb]
	\centering
	\subfloat[]{\includegraphics[width=1.7in]{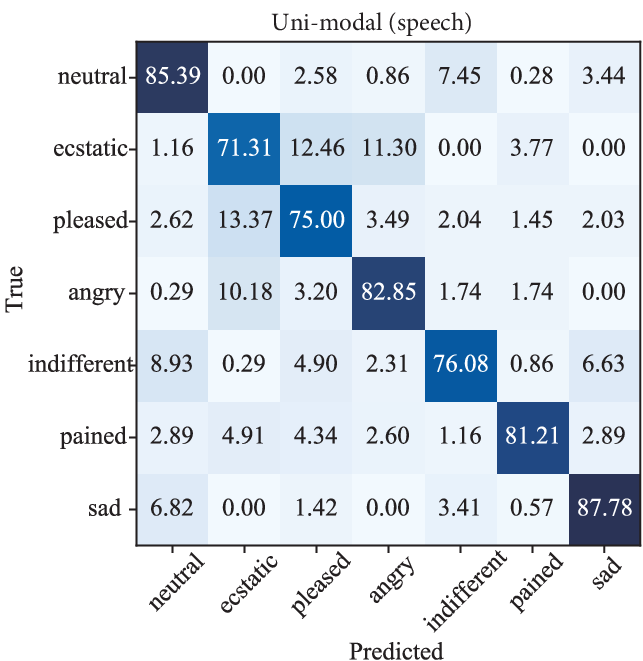}}
	\hfil
	\subfloat[]{\includegraphics[width=1.7in]{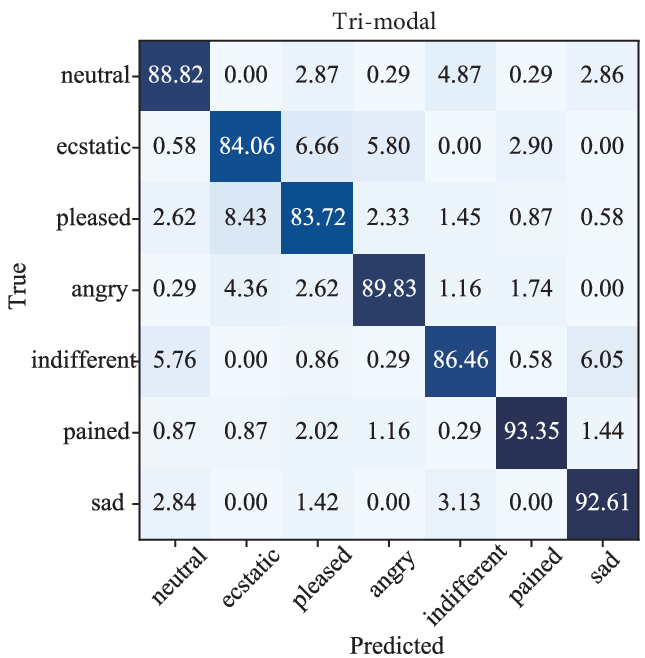}}
	\caption{Confusion matrix (in \%) for (a) the uni-modal scenario using speech and (b) the best-performing tri-modal scenario.}
	\label{Fig:cm}
\end{figure}

The confusion matrices of mono-speech and the tri-modal scenarios are displayed in FIG.~\ref{Fig:cm}. Two key confusion patterns appeared in the mono-speech case. First, the ecstatic and angry states were confused to each other, a finding consistent with several studies \cite{borchert2005, zhou2022multi}. Another common phenomenon is between the ecstatic and pleased state, which caused lower accuracy for both emotions. In the case of indifference, another emotion with low recognition accuracy in speech-only mode, misclassifications were distributed almost evenly across two moderate emotions (pleased and sad) as well as neutral. However, these main confusion problems were relieved in the tri-modal case.

\subsubsection{\label{subsec:4:2} Emotion recognition based on the estimated data}
The results of emotion recognition using estimated physiological data are shown in TABLE~\ref{tab:results_2}. In the uni-modal scenario, the predicted articularoy movements yielded the lowest accuracy, only 37.37\%, which was much lower than the accuracy obtained using real EMA. This may be caused by the poor performance of the inversion method, which obtained an average Pearson correlation coefficients of only 0.6406 between the predicted and ground-truth data. In the bi-modal scenario, combining speech with two other predicted modalities yielded better results than uni-modal speech. When three modalities were combined, the recognition accuracy was improved over all combinations in the bi-modal scenario, by 1.36\% over the highest bimodal recognition accuracy and by 2.72\% over the highest uni-modal recognition accuracy. This indicates that although the predicted articulatory data contain limited information, they still provide a small amount of information that is not available in the other two modalities.

\begin{table}
	\caption{Comparison result of uni-modal, bi-modal, tri-modal emotion recognition based on estimated physiological information. '\checkmark' means which modality is used. \label{tab:results_2}}
	\begin{center}
		\begin{tabular}{c|c|c|c}
			\hline
			\multirow{3}{*}{\textbf{Speech}} & \textbf{Excitation} & \textbf{Articulatory} & \multirow{3}{*}{\textbf{Accuracy}(\%)}\\
			& \textbf{information} & \textbf{movements}\\
			& \textbf{(estimated)} & \textbf{(estimated)}\\
			\hline
			\checkmark & & & 79.97\\
			& \checkmark & & 78.41\\
			& & \checkmark & 37.37\\
			\checkmark & \checkmark & & 81.33\\
			\checkmark & & \checkmark & 81.21\\
			& \checkmark & \checkmark & 80.51\\
			\checkmark & \checkmark & \checkmark & \textbf{82.69}\\
			\hline
		\end{tabular}
	\end{center}
\end{table}

\section{CONCLUSION}
In this paper, we conducted a preliminary study on the role of speech production-related physiological information in SER, using a combination of speech, phonation excitation information and articulatory movements. We introduced the STEM-$\rm{E^2}$VA dataset, which includes audio, EGG, and EMA data. Experimental results on STEM-$\rm{E^2}$VA confirmed the complementary value of excitation information and articulatory movements. To assess the feasibility of using such physiological information in real-world SER systems without specialized equipment, we also evaluated emotion recognition using predicted excitation information and articulatory movements from speech. Although the combination of speech and predicted physiological data improved recognition by only 2.72\% over mono-speech, the results indicate its potential for practical application. In future work, we aim to extract more salient features, optimize feature fusion techniques, and enhance inversion methods.

\bibliographystyle{IEEEtran}
\bibliography{IEEEabrv, mybibfile}

\end{document}